\RequirePackage{pdf14}

\documentclass{statsoc}
\usepackage[a4paper,margin=0.8in,bottom=1in,top=0.9in]{geometry}

\usepackage{graphicx}
\usepackage{subfigure}

\usepackage{fancyhdr,amsmath,amssymb}
\usepackage{etoolbox}
\makeatletter
\patchcmd{\@makecaption}
 {\parbox}
  {\advance\@tempdima-\fontdimen2} % decrease the width!
  {}{}
\makeatother

\renewcommand\theenumi{}

\usepackage{hyperref}
\usepackage{booktabs}
\usepackage{tikz}
\usetikzlibrary{positioning,shapes,shadows,arrows}
\usepackage{scrextend}
\title[Modelling contraceptive behavior in India via sequential logits]{Bayesian semiparametric modelling of contraceptive behavior \\ in India via sequential logistic regressions}

\author{Tommaso Rigon,}
\address{Bocconi University, Department of Decision Sciences, Via Roentgen 1, Milan, Italy}
\email{tommaso.rigon@phd.unibocconi.it}

\author{Daniele Durante,}
\address{University of Padova, Italy}

\author[Rigon, Durante and Torelli]{Nicola Torelli}
\address{University of Trieste, Italy}
\begin{document}

\begin{abstract}
Family planning has been characterized by highly different strategic programs  in India, including method-specific contraceptive targets, coercive sterilization, and more recent target-free approaches. These major changes in  family planning policies over time have motivated a considerable interest  towards assessing the effectiveness of the different programs, while understanding which subsets of the population have not been properly addressed. Current studies consider specific aspects of the above policies, including, for example, the factors associated with the choice of alternative contraceptive methods other than sterilization, for  women using contraceptives. Although these analyses produce relevant insights, they fail to provide a global overview of the different family planning policies, and the determinants underlying the contraceptive choices. Motivated by this consideration, we propose a Bayesian semiparametric model relying on a reparameterization of the multinomial probability mass function via a set of conditional Bernoulli choices. The sequential binary structure is defined to be consistent with the current family planning policies in India, and coherent with a reasonable process characterizing the contraceptive choices. This combination of flexible representations and careful reparameterizations  allows a broader and interpretable overview of the different policies and contraceptive preferences in India, within a single model. 
\end{abstract}
\keywords{Bayesian Inference, Contraceptive Method, Mixture Model, Penalized Splines, P\`olya-Gamma, Sequential Logistic Regression.}

\section{Introduction}\label{sec_1}
The rapid population growth in developing countries is a key topic in demographic  research, having immediate consequences on the increasing demand for social services, rising unemployment rates, and reduced standard of living. Although there is still a debated literature concerning the long-term effects of the over-population on the socio-economic growth \citep[e.g.][]{Blo:2012}, in the view of the developing countries, limited resources and the rapid population growth are important barriers  for the short term development process, and a main concern for governments policies.  This is particularly true in India, where the coexistence of early marriages, high poverty rates, illiteracy, and decline in infant mortality, favored a population growth rate of $1.4\%$---double than China's $0.7\%$ \citep{Blo:2012}---leading to an unsustainable population which is expected to reach $1.4$ billions over the next quarter century. 

Although India has been the first nation to introduce an official family planning program in 1951 \citep{Pac:2004, Pac:2014}, the broader access to welfare services under the clinic-based approach during  the first and second Five Year Plans in the 1950's, the subsequent focus on method-specific contraceptive targets in the mid-1960's, and the coercive sterilization programs in the 1970's and in early-1980's, failed to control the rapid population growth properly. Target-free contraceptive services, accounting for the different reproductive health needs of the population, were later promoted after the 1994  International Conference on Population and Developments \citep{Pac:2004, Pac:2014}. However,  skepticism remains about  the effectiveness of such services in  increasing contraceptive prevalence, and in stimulating a broader access to modern temporary methods, different than sterilization   \citep[e.g.][]{sav:1999}. In fact, there is evidence that the preference for permanent contraceptive practices is  still dominant compared to reversible methods, and that most of the services provided by the public programs relate to sterilization \citep{Pac:2004, Pac:2014}. Conversely, there is a growing effort by the private sector aimed at providing reproductive health services associated with reversible contraceptive methods  \citep{Pac:2004, Pac:2014}. Refer to \cite{Har:2007}, and \cite{Cha:2014} for additional details and timelines of family planning programs in India. 

The above differences in the family planning services, combined with the marked socio-demographic inequalities characterizing the population in India, have increased the emphasis on the availability of strategic datasets and statistical models to evaluate the family planning policies, and to identify which subsets of the population have not been properly addressed. As a result, detailed  surveys, such as the India Human Development Survey II (IHDS-II) and the National Family Health Survey (NFHS), have been recently conducted, motivating an increasing interest on the determinants of contraceptive choice in the light of the current policies. In this contribution we explore the IHDS-II survey data at \url{http://www.icpsr.umich.edu/icpsrweb/ICPSR/studies/36151},  to provide a flexible overview on the different policies and preferences, within a single model. The dataset is described in detail below.

\subsection{India Human Development Survey-II (IHDS-II)}\label{sec_11}
Contraceptives play a relevant role in family planning, and the widely accepted association between the contraceptive prevalence rates and the total fertility rates \citep[e.g.][]{Mau:1988} drives extensive demographic research on developing countries. Consistent with these interests, we focus our analysis on the IHDS-II 2011-2012 module devoted to ever-married women, which provides information about contraceptive choices, along with other socio-demographic variables.

The IHDS-II 2011-2012 is a nationally representative multi-topic survey conducted on $42152$ households  over $33$ States of India \citep{Des:2015}. The survey arises from the collaboration between the University of Maryland and the National Council of Applied Economic Research (NCAER) in New Delhi, and is divided in different modules aimed at monitoring a wide range of socio-economic behaviors.  Eligible units in our analysis are women aged $15-49$ who have been married at least once in their life, although their current marital status may not be married, and who were not pregnant at the moment of the interview. Recalling our research interests from Section \ref{sec_1}, the response in our study is a qualitative variable having four mutually exclusive outcomes:
\begin{enumerate}
\item{{\bf [1]} \ \texttt{no contraceptive}: no contraceptive method is used.}
\item{{\bf [2]} \ \texttt{sterilization}: the woman or her partner underwent sterilization or hysterectomy.}
\item{{\bf [3]} \ \texttt{natural methods}: either withdraw or periodic abstinence is chosen.}
\item{{\bf [4]} \ \texttt{modern methods}: the woman or her partner use modern methods (e.g. oral pill, condom).}
\end{enumerate}

A small set of women declared to use contraceptives different than those listed above. Since we do not have information about these alternative methods, we held them out from our analysis. Consistent with this choice, we also do not consider women for whom the information on contraceptive preference is not observed. This preliminary pre-processing provides a final sample size of $n=30524$. Although finer classifications of contraceptive methods could be considered, the four categories listed above are those of main interest in family planning policies   \citep[e.g.][]{Pac:2004}, and are commonly considered in statistical modelling of contraceptive behavior in India   \citep[e.g.][]{Deo:2014}.

In selecting the covariates of interest we leverage instead recent  evidences from statistical analyses on contraceptive behavior in India  \citep[e.g.][]{Deo:2014},  and more general discussions on the relevant factors underlying contraceptive preferences, provided by social science studies  \citep[e.g.][]{Pac:2004}. More specifically, we consider the \texttt{AGE} information, a binary variable \texttt{AREA}  indicating  whether the woman lives in \texttt{urban} or \texttt{rural} areas, a four level \texttt{RELIGION} factor, encoding \texttt{hindu}, \texttt{muslim}, \texttt{christian}, or \texttt{other} religions, a categorical variable \texttt{EDUCATION} classifying women according to \texttt{no} education, \texttt{low} education, \texttt{intermediate} education or \texttt{high} education, and a grouping variable \texttt{CHILD} for women having \texttt{no child}, \texttt{one child} or \texttt{more than one child}, respectively. We also exploit the information on the \texttt{STATE} of residence to define the hierarchy in our model via State-specific effects. These quantities are also of interest for inference, in providing information on across-State differences in contraceptive preferences, after controlling for the socio-demographic covariates. Although we could consider additional covariates, our main goal is to disambiguate and interpret the effect of the most studied variables at the different steps of the contraceptive choices.

As discussed in Section \ref{sec_21}, we propose to analyze the above data under a statistical model which relies on a set of sequential binary comparisons among subsets of contraceptive choices. This focus is explicitly motivated by the current family planning policies in India discussed in Section \ref{sec_1}, and is also coherent with a reasonable decision process underlying contraceptive preferences, thereby providing more general and interpretable inference compared to classical analyses based on standard parameterizations of the multinomial logistic regression. Motivated by the marked socio-demographic differences characterizing the population in India, we also improve flexibility in modelling the covariates effects at the different steps of the sequential binary choices under a Bayesian semiparametric formulation outlined in Section  \ref{sec_22}. Beside facilitating flexible and interpretable inference, the proposed methods are also associated with simple algorithms for inference; see Section  \ref{sec_3}. As carefully outlined in Section \ref{sec_4}, this combination of flexible representations for the covariates effects, and careful reparameterizations of the multinomial likelihood for the contraceptive preference data, provide relevant and interpretable insights for policy makers, while improving predictive performance. These results, and the source code to reproduce them, are made available at \url{https://github.com/tommasorigon/India-SequentiaLogit} along with an interactive {\bf Shiny} application. Concluding remarks  are provided in Section  \ref{sec_5}. 

\section{Bayesian semiparametric modelling of contraceptive preferences}\label{sec_2}
As discussed in Section \ref{sec_1}, we reparameterize the multinomial probability mass function for the contraceptive methods via conditional Bernoulli choices for subsets of contraceptives, and provide inference on the socio-demographic factors underlying these sequential binary comparisons via a Bayesian semiparametric representation for the covariates effects. Sections \ref{sec_21} and \ref{sec_22} describe these generalizations, with a specific reference to the research interests associated with the contraceptive behavior in India.

\subsection{Model formulation via sequential Bernoulli choices}\label{sec_21}
Let ${\bf y}_{ij} = (y_{ij1}, y_{ij2}, y_{ij3},y_{ij4})$, denote the vector of binary variables encoding the contraceptive choice of woman $j=1, \ldots, n_i$ in State $i=1, \ldots, 33$, with:
\begin{enumerate}
\item{{\bf [1]} \ ${\bf y}_{ij} = (1, 0, 0,0)$ if woman $j$ in State $i$ and her partner use no contraceptive methods.}
\item{{\bf [2]} \ ${\bf y}_{ij} = (0, 1, 0,0)$ if woman $j$ in State $i$ or her partner underwent sterilization.}
\item{{\bf [3]} \ ${\bf y}_{ij} = (0, 0, 1,0)$ if woman $j$ in State $i$ and her partner use natural methods.}
\item{{\bf [4]} \ ${\bf y}_{ij} = (0, 0, 0,1)$ if woman $j$ in State $i$ and her partner use modern methods.}
\end{enumerate}
Following standard procedures in statistical modelling of categorical response data we let
\begin{eqnarray}
\label{eq1.1}
{\bf y}_{ij} \mid \boldsymbol\pi_{ij} &\sim& \mbox{Multinom}(1,\boldsymbol\pi_{ij}), \qquad \boldsymbol\pi_{ij} = (\pi_{ij1},\pi_{ij2},\pi_{ij3},\pi_{ij4}), 
\end{eqnarray}
independently for each State $ i=1,\ldots,33$ and woman $j=1,\ldots,n_i$, where $\pi_{ijr} \in (0,1)$ indicates the probability that woman $j$ in State $i$ adopts the contraceptive behavior $r$, listed in Section \ref{sec_11}.

Under the usual specification of the multinomial regression \citep[e.g.][]{Agr:2007}, one can pair each contraceptive choice with a reference category---characterizing, for example, no use of contraceptives---and then model the log-odds of every pair as a function of the covariates.  Although this is a common specification \citep[e.g.][]{Jaya:2009, Hus:2013, Ram:2014}, the resulting inference and conclusions are confined to the pairwise comparisons with the selected reference category, thereby ruling out the possibility to learn---within a single statistical model---the direct effect of the covariates on other log-odds of potential interest for policy makers, including conditional choices among subsets of similar contraceptive methods. Indeed, these conditional quantities are typically of interest, and are implicitly involved in the statistical analysis of the contraceptive behavior.  For example,  \cite{Deo:2014} rely on a multinomial logistic regression, applied to the subset of women currently using contraceptives, to disentangle the socio-demographic factors associated with the preference of modern and traditional contraceptive methods compared to sterilization.  \cite{Mis:2014} consider instead different logistic regressions to learn the covariates effects on the preference of a specific contraceptive method---or a subset of methods---for different combinations of interest. 

\tikzstyle{abstract}=[rectangle, draw=black, rounded corners, anchor=north, text=black, text width=4.5cm]
\tikzstyle{myarrow}=[<-, >=open triangle 90, thick]
\tikzstyle{line}=[-, thick]
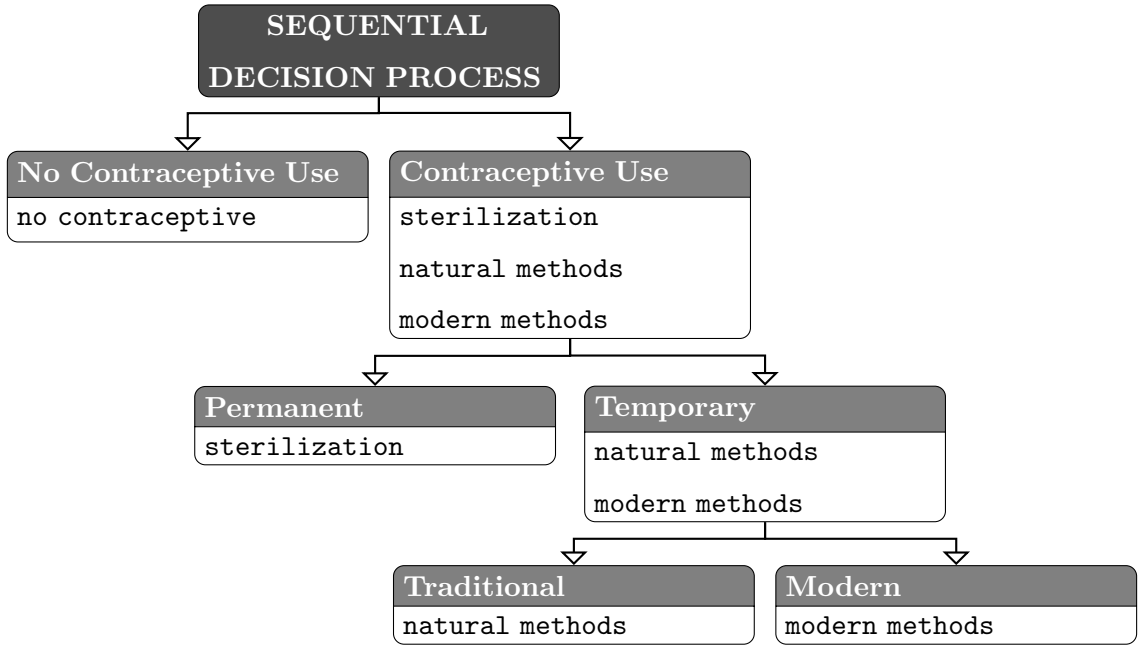
\begin{figure}[t]        
\begin{center}
\resizebox{15cm}{!}{
\begin{tikzpicture}[node distance=2cm]
    \node(Item) [abstract, rectangle split, rectangle split parts=1,minimum size=19pt,rectangle split part fill={black!70,white},]
        {
            \textcolor{white}{\ \ \ \ \ \  \textbf{SEQUENTIAL \\ DECISION PROCESS}}
            
               };
    \node (AuxNode01) [text width=4cm, below=of Item] {};
    \node (Component) [abstract, rectangle split, rectangle split parts=2,rectangle split part fill={gray,white}, right=of AuxNode01,xshift=-4cm,yshift=5]
        {
            \textcolor{white}{\textbf{Contraceptive Use}}
            \nodepart{second} \texttt{sterilization} \\  \texttt{natural methods} \\  \texttt{modern methods}
        };
    \node (System) [abstract, rectangle split, rectangle split parts=2,rectangle split part fill={gray,white}, left=of AuxNode01,xshift=4cm,yshift=23.2]
        {
            \textcolor{white}{\textbf{No Contraceptive Use}}
            \nodepart{second} \texttt{no contraceptive}
        };
    \node (AuxNode02) [text width=0.5cm, below=of Component] {};
    \node (Sensor) [abstract, rectangle split, rectangle split parts=2,rectangle split part fill={gray,white}, right=of AuxNode02,xshift=-2.2cm,yshift=17.3]
        {
            \textcolor{white}{\textbf{Temporary}}
            \nodepart{second}  \texttt{natural methods} \\ \texttt{modern methods} 
        };
    \node (Part) [abstract, rectangle split, rectangle split parts=2,rectangle split part fill={gray,white}, left=of AuxNode02,xshift=2.2cm,yshift=28]
        {
            \textcolor{white}{\textbf{Permanent}}
            \nodepart{second} \texttt{sterilization} 
        };
        
    \node (AuxNode03) [below=of Sensor] {};
    \node (Pressure) [abstract, rectangle split, rectangle split parts=2,rectangle split part fill={gray,white}, right=of AuxNode03, xshift=-2cm,yshift=29.5]
        {
            \textcolor{white}{\textbf{Modern}}
            \nodepart{second}   \texttt{modern methods} 
        };
    \node (Temperature) [abstract, rectangle split, rectangle split parts=2,rectangle split part fill={gray,white}, left=of AuxNode03, xshift=2cm,yshift=29.5]
        {
            \textcolor{white}{\textbf{Traditional}}
            \nodepart{second}  \texttt{natural methods}
        };

       \draw[myarrow] (Component.north) -- ++ (0,0.5) -| (Item.south);
        \draw[myarrow] (System.north) -- +(0,0.5) -| (Item.south);
        
          \draw[myarrow] (Sensor.north) -- ++ (0,0.4) -| (Component.south);
        \draw[myarrow] (Part.north) -- +(0,0.4) -| (Component.south);
        
          \draw[myarrow] (Pressure.north) -- ++ (0,0.35) -| (Sensor.south);
        \draw[myarrow] (Temperature.north) -- +(0,0.35) -| (Sensor.south);
   \end{tikzpicture}}
   \vspace{-18pt}
\caption{Representation of the sequential process underlying the contraceptive behavior.}\label{F_decision}
\end{center}
\end{figure}

The above contributions arguably rely on more interpretable parameterizations and comparisons, which can be reformulated depending on the research interests. For instance, when the main goal is to evaluate policies aimed at increasing contraceptive prevalence under the current target-free `cafeteria' approach in India \citep{Pac:2004, Pac:2014},  it is arguably more coherent and of direct interest to study the socio-demographic factors underlying the binary decision to use or not contraceptives, rather than modeling the log-odds of each contraceptive method with no contraceptive usage as reference category. 

Motivated by these considerations, we rely on a reparameterization of the multinomial probability mass function via a sequence of Bernoulli choices between increasingly nested subsets of contraceptive methods. The binary tree structure of interest is represented in Figure~\ref{F_decision}, and is defined to characterize a reasonable decision process underlying the contraceptive choices. In particular, as shown in Figure~\ref{F_decision}, this decision process starts with the choice of using or not contraceptives. If a contraceptive method is chosen, the next step requires deciding between permanent or temporary contraceptives.  Finally---in case a temporary method is preferred---the choice is between traditional or modern methods. Our overarching focus is to infer the socio-demographic effects underlying each step of this decision process.

Although other nested decision processes can be devised, and our inference procedures can be easily adapted to different binary tree structures,  the sequential mechanism in Figure~\ref{F_decision} allows interpretable inference on dependence structures of direct interest in the light of the current India family programs, within a single statistical model. In particular, disentangling---as a first step---the socio-demographic factors associated with the contraceptive prevalence, is coherent with the current `cafeteria' approach in India \citep{Pac:2004, Pac:2014}, and can provide relevant insights on which subsets of the population in India have not been currently addressed. Indeed, the contraceptive prevalence rate is one of the main performance indicators for family planning \citep[e.g.][]{Alk:2013}, and several analyses focus directly  on this binary comparison, pooling the different contraceptive methods \citep[e.g.][]{Mcn:2003,Dha:2004,  DeoDias:2014, Haq:2015}. 

The second step recalls, instead, the model proposed by \cite{Deo:2014}, and is motivated by the fundamental interest underlying the factors associated with the ongoing dominance of sterilization compared to alternative methods promoted by target-free programs in India. Consistent with \cite{Deo:2014}, this analysis focuses on the subsets of women currently using a contraceptive method, in order to  isolate the inference from the effects of use versus non-use of contraceptives. However, differently from \cite{Deo:2014}---who consider a multinomial logistic regression for sterilization, traditional, and modern methods---we first study the choice between sterilization and the alternative temporary methods, and then focus on the decision between traditional and modern methods, for the subset of non-sterilized women. This arguably allows more direct inference---in the second step---on the socio-demographic factors underlying the general preference for sterilization, which is of main interest when to goal is to understand the reasons for the failure of the family planning policies in motivating broader access to  temporary methods  \citep[e.g.][]{sav:1999, Pac:2004, Pac:2014}. 

Finally, as discussed in Section \ref{sec_1}, there is a growing effort by the private sector in India towards addressing reproductive health needs other than sterilization, with a focus on modern reversible methods  \citep[e.g.][]{Pac:2004, Pac:2014}. Consistent with this, the last step  in Figure~\ref{F_decision} focuses on the choice between natural and modern methods, for those women currently using reversible contraceptives. This group arguably represents the segment of more direct interest for  the private sector \citep[e.g.][]{Pac:2004, Pac:2014}, and learning the determinants underlying the preference for natural methods instead of modern ones, can provide key insights for the private sector to assess and improve targeting strategies.

As a consequence of the above sequential process, the explicit focus of inference is  not directly on the vector of marginal probabilities $\boldsymbol\pi_{ij} = (\pi_{ij1},\pi_{ij2},\pi_{ij3},\pi_{ij4})$ for the different contraceptive behaviors, but on the conditional probabilities $\boldsymbol \rho_{ij} = (\rho_{ij1},\rho_{ij2},\rho_{ij3}, \rho_{ij4})$, defined as
\begin{align}
\label{eq1.2}
\rho_{ij1} = \sum_{r=2}^4\pi_{ijr}, \quad \ \rho_{ij2} = \frac{\sum_{r=3}^4\pi_{ijr}}{\sum_{r=2}^4\pi_{ijr}}, \quad \ \rho_{ij3} = \frac{\pi_{ij4}}{\sum_{r=3}^4\pi_{ijr}}, \quad \ \rho_{ij4} =\frac{\pi_{ij3}}{\sum_{r=3}^4\pi_{ijr}}=1-\rho_{ij3},
\end{align}
for every State $ i=1,\ldots,33$ and woman $j=1,\ldots,n_i$. This reparametrization facilitates inference on the conditional probabilities characterizing the sequential process in Figure \ref{F_decision}. In fact, $\rho_{ij1}$ represents the probability of using contraceptives, whereas $\rho_{ij2}$ denotes the conditional probability of considering a reversible method, given the decision of using any contraceptive. Finally---following the sequential decision process in Figure \ref{F_decision}---the parameters $\rho_{ij3}$ and $\rho_{ij4}$ measure the probability of using a modern or traditional contraceptive method, respectively,  conditionally on the adoption of temporary methods. 

Differently from \cite{Deo:2014}, equation \eqref{eq1.2} allows direct modelling of the entire range of contraceptive behaviors, instead of just a subset of them, thereby providing more general inference and prediction, which is of interest for a wider spectrum of family planning policies. Moreover, inference is performed within a single statistical model based on a reparameterization of the multinomial probability mass function, providing coherent methods for prediction  and uncertainty quantification. This is not the case in  \cite{Mis:2014} whose conditional inference on combinations of contraceptive methods, cannot be recast within a single statistical model. Conversely, under the proposed reparameterization in  equation \eqref{eq1.2}, the multinomial probability mass function $\mbox{pr}({\bf y}_{ij})= \pi_{ij1}^{y_{ij1}}\pi_{ij2}^{y_{ij2}}\pi_{ij3}^{y_{ij3}}\pi_{ij4}^{y_{ij4}}$  for each statistical unit  can be formally re-written   as the product of Bernoulli probability mass functions for the sequential binary comparisons in Figure \ref{F_decision}, obtaining:
\begin{eqnarray}
\mbox{pr}({\bf y}_{ij})=[\rho_{ij1}^{y_{ij2}+y_{ij3}+y_{ij4}}(1-\rho_{ij1})^{y_{ij1}}]\cdot[\rho_{ij2}^{y_{ij3}+y_{ij4}}(1-\rho_{ij2})^{y_{ij2}}]\cdot[\rho_{ij3}^{y_{ij4}}(1-\rho_{ij3})^{y_{ij3}}],
\label{repa}
\end{eqnarray}
 where the first Bernoulli variable characterizes the choice between no use $(y_{ij1}=1, y_{ij2}+y_{ij3}+y_{ij4}=0)$ and use $(y_{ij1}=0, y_{ij2}+y_{ij3}+y_{ij4}=1)$ of contraceptives, whereas the second represents the decision among sterilization $(y_{ij2}=1, y_{ij3}+y_{ij4}=0)$ and reversible methods $(y_{ij2}=0, y_{ij3}+y_{ij4}=1)$, for those women using  contraceptives. Finally, the third Bernoulli variable focuses on the choice between natural ($y_{ij3}=1, y_{ij4}=0$) and modern methods ($y_{ij3}=0, y_{ij4}=1$), for the women currently using temporary contraceptives. Refer to \cite{Tutz:1991} for an overview of the sequential logistic regression.

Consistent with our goals, and motivated by results in \eqref{repa}, we aim to learn the socio-demographic factors underlying the sequential binary choices in Figure \ref{F_decision}, via a logistic regression for each  conditional probability in equation \eqref{eq1.2}. To accomplish this goal accurately, we seek a flexible representation for the relation between the conditional probabilities and the covariates, which allows coherent uncertainty quantification, efficient inference,  and possible inclusion of prior information. Consistent with these aims, we rely on a Bayesian semiparametric approach, which provides an appealing direction in hierarchical demographic models characterized by nested structures---such as those outlined in  Figure \ref{F_decision}---and by the need to borrow information across the observed data to improve inference on parameters for which limited information is available. As discussed in Section \ref{sec_22}, this is particularly useful in flexibly modelling the functional effect of age, and the changes in contraceptive preferences across States characterized by limited data. Refer to \cite{Bij:2016}, and \cite{Eld:2016} for a careful discussion on the benefits of Bayesian demographic inference in characterizing hierarchical and complex representations, quantifying and propagating uncertainty, borrowing of information, and incorporating possible prior knowledge. Besides these benefits, there are also practical computational advantages, allowing simple posterior inference for the parameters of interest, covering the covariates effects, along with the conditional and the marginal probabilities in \eqref{eq1.2} and \eqref{eq1.1}, respectively.

\subsection{Bayesian semiparametric logistic regressions}\label{sec_22}
Recalling the discussion in Section \ref{sec_21}, the main focus is on learning the effects of the covariates on the conditional probabilities in equation \eqref{eq1.2}, via the set of logistic regressions having additive effects
\begin{eqnarray}
\label{linpred}
\mbox{logit}( \rho_{ijk})=\log\left(\frac{\rho_{ijk}}{1 - \rho_{ijk}}\right)= \mu_{ik} + f_k(\texttt{age}_{ij}) + {\bf{x}}_{ij}^\intercal \boldsymbol\beta_k, \qquad k=1,\ldots,3,
\end{eqnarray}
where ${\bf{x}}_{ij}$ is the vector of covariates encoding the variables \texttt{AREA}, \texttt{RELIGION}, \texttt{EDUCATION}, and \texttt{CHILD} into dummy indicators, with associated vector of coefficients $\boldsymbol\beta_k$. One category for each qualitative covariate is left out in \eqref{linpred}, and considered as baseline, for identifiability. Hence the  vector of coefficients $\boldsymbol\beta_k$ represents incremental effects with respect to the baseline categories. Representation \eqref{linpred} consists of a State-specific intercept $\mu_{ik}$, a functional effect of the variable \texttt{AGE}, and a set of dummy variables to learn changes in the conditional probabilities of interest with the categories of the qualitative variables \texttt{AREA}, \texttt{RELIGION}, \texttt{EDUCATION}, and \texttt{CHILD}. To avoid identifiability issues, the first State-specific intercept $\mu_{1k}$,  corresponding to the most populated State in our sample---Uttar Pradesh---has been fixed to zero, and therefore $\mu_{2k},\dots,\mu_{33k}$ measure incremental effects with respect to this baseline State. Although identifiability could be incorporated also via restrictions on the prior or via post-processing \citep[e.g.][]{Li:2011}, we prefer to enforce  identifiability directly in the likelihood to facilitate interpretation, and possible implementation in non-Bayesian inference settings.

To be more specific, the three logistic regressions of interest are
\begin{equation}
  \begin{aligned}
\mbox{{\bf Usage:}} && \mbox{pr}(\mbox{contraceptive use})= \rho_{ij1},  \ \ \ \mbox{logit}( \rho_{ij1})=\mu_{i1} + f_1(\texttt{age}_{ij}) + {\bf{x}}_{ij}^\intercal \boldsymbol\beta_1, \ \ \ \ \ \\
\mbox{{\bf Reversibility:}}    && \mbox{pr}(\mbox{temporary} \mid \mbox{contraceptive use})= \rho_{ij2}, \ \ \  \mbox{logit}( \rho_{ij2})=\mu_{i2} + f_2(\texttt{age}_{ij}) + {\bf{x}}_{ij}^\intercal\boldsymbol\beta_2, \ \ \ \ \ \\
\mbox{{\bf Method:}}   &&  \mbox{pr}(\mbox{modern} \mid \mbox{temporary})= \rho_{ij3}, \  \ \ \mbox{logit}( \rho_{ij3})=\mu_{i3} + f_3(\texttt{age}_{ij}) + {\bf{x}}_{ij}^\intercal \boldsymbol\beta_3,  \ \ \ \ \
\label{seq_logi}
\end{aligned}
\end{equation}
and our focus is on providing flexible Bayesian inference on the parameters in \eqref{seq_logi}. We consider the logistic link, instead of other alternatives---such as the probit---since it provides a common and more interpretable choice in these types of analyses, while representing the canonical link in the exponential family representation of the Bernoulli random variable.

In modeling the State effects $\mu_{ik}$, $i=2, \ldots, 33$, $k=1, \ldots, 3$---or similar community-level covariates---current studies consider either fixed parameters within a classical generalized linear model framework \citep[e.g.][]{Jaya:2009, Rai:2013, Ram:2014}, or include Gaussian random effects under a multilevel representation \citep[e.g.][]{Mcn:2003, Dha:2004, Deo:2014, DeoDias:2014}. Although this generalization is appealing in accounting for the hierarchical structure of the data, the resulting borrowing of information and uncertainty quantification can be quite sensitive to departures from the normality assumption \citep[e.g.][]{Dun:2010}, which are arguably expected in our study. Indeed, recalling our application, it is reasonable to expect  States with common unobserved characteristics such as cultural acceptance, accessibility, or women social condition to have a comparable effect on the decision process underlying the contraceptive behaviors. Moreover, such effects may vary substantially between groups of States due to the socio-economic interstate differences in India  \citep[e.g.][]{Bal:2010}. Hence, assuming a common Gaussian  distribution may force the State-specific random effects  to over-shrink around the population mean, while  ruling out possible clustering among States. Consistent with this discussion, we consider a flexible representation for the prior distribution of the States-specific effects, and incorporate clustering by replacing the common Gaussian assumption, with a location mixture of Gaussians
 \begin{eqnarray}
\mu_{ik} \mid \Pi_k \sim \Pi_k, \ i=2, \ldots, 33, \quad \mbox{with } \Pi_k =\sum_{h=1}^H  \nu_{hk} \mbox{N}(\bar{\mu}_{hk},\sigma_k^{2}),  \quad \mbox{independently for } k=1, \ldots, 3, 
\label{eq5}
\end{eqnarray}
having priors for the mixing probabilities $(\nu_{1k}, \ldots, \nu_{Hk})$, and kernel parameters $\bar{\mu}_{1k}, \ldots, \bar{\mu}_{Hk},$ and $\sigma_{k}^{2}$
 \begin{eqnarray}
(\nu_{1k}, \ldots, \nu_{Hk}) \sim \mbox{Dirich}(1/H, \ldots, 1/H), \  \bar{\mu}_{hk} \sim \mbox{N}(0, \sigma_{\mu k}^2), \ h=1, \ldots, H, \ \sigma_{k}^{-2}=\tau_k \sim \mbox{Ga}(a_{\tau_k}, b_{\tau_k}),
\label{eq6}
\end{eqnarray}
for every $k=1, \ldots, 3$. 

A key result in \eqref{eq5} is that the mixture representation favors ties among State-specific effects, with States sharing the same cluster having the same Gaussian prior. Specifically let $G_{ik}$ denote the cluster indicator of State $i$ in the $k$th sequential logistic regression---with $\mbox{pr}(G_{ik}=h)=\nu_{hk}$---the mixture of Gaussians prior favors clustering effects among the States, with $(\mu_{ik} \mid G_{ik}=h) \sim \mbox{N}(\bar{\mu}_{hk},\sigma_k^{2})$, for each $i=2,\ldots,33$, and $k=1, \ldots,3$. This property is particularly useful in our application, favoring States with common unobserved characteristics to share the same Gaussian prior.  Note also that, in \eqref{eq6} the mixing probabilities $(\nu_{1k}, \ldots, \nu_{Hk})$ have a Dirichlet prior with parameters $(1/H, \ldots, 1/H)$. This choice is motivated by recent theoretical results on recovering the true number of components in Gaussian mixture models \citep{Rou:2011}. When all the mixture components---except one---are empty, representation \eqref{eq5} reduces to a common Gaussian prior for all the State-specific effects, so that classical multilevel models for contraceptive preferences are special cases of our representation.

\begin{figure}
\centering
\includegraphics[height=5.1cm, width=17cm]{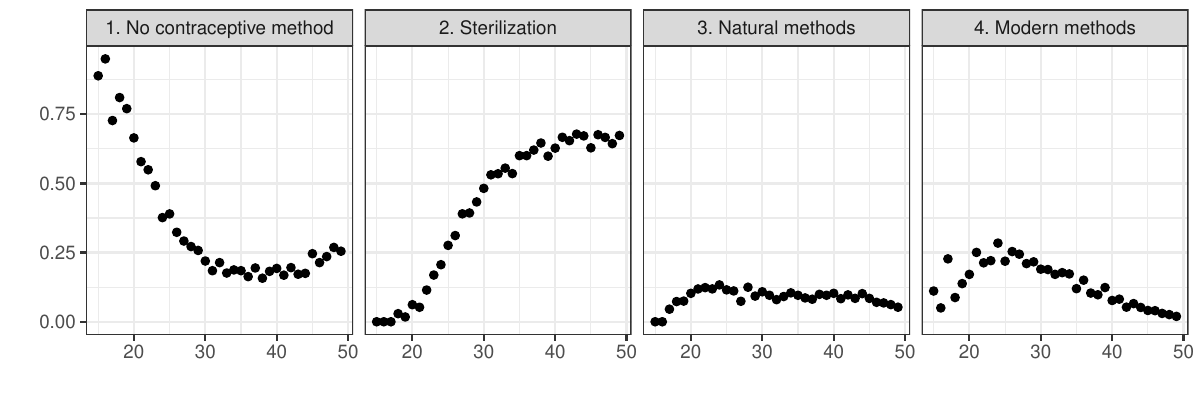}
\caption{\label{predictor}{\footnotesize{Observed relative frequency of contraceptive users at different ages displayed for every contraceptive method under analysis: no contraceptive, sterilization (female and male), natural methods (including withdraw and periodic abstinence),  modern methods (condom, oral pill, copper IUD and others).}}}
\end{figure}

We seek a similar flexibility also in characterizing the functional effect of the variable \texttt{AGE} in \eqref{linpred}, away from classical linear parametric representations \citep[e.g.][]{Hus:2013}. In fact, as shown in Figure \ref{predictor}, such assumption may be overly-restrictive in our application, thereby affecting the quality of inference.  Introducing specific parameters for classes of age as in \cite{ Mcn:2003,  Dha:2004, Rai:2013, Ram:2014,  Deo:2014,  DeoDias:2014, Mis:2014}  improves flexibility, but questions remain on the number and location of the thresholds, which may lead to substantially different results. Motivated by this consideration, we avoid a pre-specified functional form for  $f_k(\cdot)$, $k=1, \ldots, 3$, and model the unknown functions $f_k(\cdot)$ via a flexible  linear combination of B-spline basis functions $\mathcal{B}_m(\cdot)$, $m=1, \ldots, M$, obtaining
\begin{eqnarray}
\label{eq:splines}
f_k(\texttt{age}_{ij})  = \sum_{m=1}^{M} \gamma_{mk}\mathcal{B}_m(\texttt{age}_{ij}), \quad \mbox{independently for } k=1, \ldots, 3,
\label{eq4}
\end{eqnarray}
where $M = n_\texttt{knots} - 4$ is the total number of basis functions, $n_\texttt{knots}$ denotes the total number of knots characterizing the B-spline basis, and $\boldsymbol{\gamma}_k = (\gamma_{1k},\dots,\gamma_{Mk})$ represents the vector of real coefficients governing the linear combination of the B-spline basis which characterizes $f_k(\texttt{age}_{ij})$.

Equation \eqref{eq4} defines a flexible representation for the functional effect of the  variable \texttt{AGE} in \eqref{linpred}, and has been successfully considered in different demographic applications \citep[e.g.][]{Mcnei:1977}, beyond contraceptive studies. However, it requires the choice of the number and location of the knots, with poor choices leading either to possible bias or overfitting. To overcome this difficulty, we follow \citet{Eil:1996} by relying on a sufficiently large number of equally spaced knots, and impose the penalty $\lambda_k \sum_{m=3}^M(\gamma_{mk} - 2\gamma_{(m-1) k} + \gamma_{(m-2) k})^2$, in the log-likelihood to penalize  highly parameterized representations. This penalty forces subsequent coefficients to be similar, thus controlling the smoothness of the function $f_k(\cdot)$ by an amount which depends directly on the smoothing parameter $\lambda_k>0$. As discussed in \citet{Lan:2004}, this penalization can be rephrased within a Bayesian framework by assuming a particular rank-deficient Gaussian priors for each set of coefficients $\boldsymbol\gamma_k$
\begin{eqnarray}
 \boldsymbol\gamma_k \mid \lambda_k \sim \mbox{N}(0, {\lambda_k^{-1}{\bf D}^+}), \quad \mbox{independently for } k=1, \ldots, 3,
\label{eq4bis}
\end{eqnarray}
where $\bf D^+$ represents the pseudo-inverse of a suitable positive semi-definite matrix $\bf{D}$, which incorporates the aforementioned quadratic penalty. To learn the smoothness of each function, we additionally define a Gamma hyperprior $\lambda_k \sim \text{Ga}(a_\lambda, b_\lambda)$, for each $k=1,\dots,3$.

To conclude our Bayesian formulation we consider a multivariate Gaussian prior for the dummy coefficients $\boldsymbol\beta_k$, $k=1, \ldots, 3$, in \eqref{linpred} by letting
\begin{eqnarray}
\boldsymbol\beta_k \sim \mbox{N}( {\bf{b}}, {\bf{B}} ), \quad \mbox{independently for } k=1, \ldots, 3,
\label{eq3}
\end{eqnarray}
where ${\bf{b}}$ and ${\bf{B}}$ are the prior mean vector and covariance matrix, respectively. Note also that, since the variables \texttt{AREA}, \texttt{RELIGION}, \texttt{EDUCATION}, and \texttt{CHILD} are qualitative, characterizing these covariates via a set of dummy indicators for each category, already provides a fully flexible specification for the additive effects of  \texttt{AREA}, \texttt{RELIGION}, \texttt{EDUCATION}, and \texttt{CHILD}.

%We generalize the above mentioned procedures in two different directions in order to develop a  flexible model-based approach to explore the determinants of contraceptive behaviors in India. Specifically we allow the functional effect of the variable age entering the predictor via a semiparametric specification through Bayesian penalized splines, and exploit the Dirichlet process (DP) priors to enforce uncertainty in the distribution of the India's States-specific random intercepts, while favoring clustering effects. Hence we are more closely related to the literature on functional data analysis with proposal covering semiparametric approaches via spline functions \citep[e.g.][]{Azz:2012} and nonparametric regression via Gaussian processes \citep{Ras:2006}. For a review on Bayesian nonparametric modelling via DP prior see \citet{Hjo:2010}.  Such models are widely used in machine learning, biostatistics and epidemiological applications, with recent growth of interest in demographic and social science settings \citep[e.g.][]{Kee:2008, Kyu:2011}. However, similar contributions analyzing the determinants of contraceptive use are lacking and researches focusing on India \citep[e.g.][]{Cha:2001,  Mcn:2003, Dha:2004, Rai:2013 } could benefit from an improved flexibility in the statistical model. 

Beside providing a statistical model which is coherent with our research interests, the three logistic regressions defined in equations \eqref{seq_logi} can be studied separately---according to factorization \eqref{repa}. Within a Bayesian framework, this property has the key computational benefit of allowing separate Markov chain Monte Carlo algorithms for posterior computation, provided that the prior distributions \eqref{eq5}--\eqref{eq3} for the parameters in equation \eqref{linpred} are specified independently for $k=1, \ldots, 3$. Although it would be possible to introduce dependence among the covariates effects also across the three logistic regressions, the increment in efficiency may be low relative to the restrictions induced by this higher-level borrowing of information. Hence, we prefer to avoid other hierarchical layers, which may unnecessarily increase model complexity and computational intractability. In fact, as we will discuss in Section 4, maintaining independence among the priors in the different logistic regressions does not affect efficiency and already provides an effective representation. The priors are also defined to  maintain full conditional conjugacy for simple implementation of the  Gibbs sampler described in Section \ref{sec_3}.

\section{Posterior computation}\label{sec_3}
Posterior computation relies on a recently developed data augmentation scheme for Bayesian logistic regression, based on P\'olya-gamma variables; see \citet{Pol:2013} for a detailed description, and \citet{Cho:2013} for theoretical results.  This approach provides a strategy for full conditional conjugate Bayesian inference exploiting the representation of the binomial likelihood---parameterized via the log-odds---as a scale mixture of Gaussians with P\'olya-gamma mixing measure. Specifically assuming a Bayesian logistic regression $y_i \sim \mbox{Bern}[\{1+\exp(-{\bf{x}}_i^\intercal \boldsymbol\beta)\}^{-1}]$, $i=1, \ldots, n$, with  $\boldsymbol\beta \sim \mbox{N}({\bf{b}},{\bf{B}})$, the Gibbs sampler relying on the P\'olya-gamma data augmentation exploits the fact that, given the P\'olya-gamma data $\omega_i \sim  \mbox{PG}(1,{\bf{x}}_i^\intercal \boldsymbol\beta)$, the contribution to the likelihood for the $i$th statistical unit is
\begin{eqnarray*}
\propto  \exp \left\{-\frac{\omega_i}{2} \left(\frac{y_i-1/2}{\omega_i}-{\bf{x}}_i^\intercal \boldsymbol\beta\right)^2 \right\}, \quad \mbox{for every } i=1, \ldots, n.
\end{eqnarray*}
Hence, leveraging the above representation based on Gaussian kernels, the resulting Gibbs sampler simply alternates between the two full conditional steps
\begin{eqnarray*}
\omega_i \mid \boldsymbol\beta, {\bf{x}}_i \sim \mbox{PG}(1,{\bf{x}}_i^\intercal \boldsymbol\beta)\quad \mbox{and} \quad \boldsymbol\beta \mid {\bf{y}},\boldsymbol\omega, {\bf{x}} \sim \mbox{N}(\boldsymbol\mu_{\beta}, \boldsymbol\Sigma_{\beta}),
\end{eqnarray*}
with $\boldsymbol\Sigma_{\beta}=({\bf{X}}^\intercal \boldsymbol\Omega {\bf{X}}+{\bf{B}}^{-1})^{-1}$, $\boldsymbol\mu_{\beta}=\boldsymbol\Sigma_{\beta}({\bf{X}}^\intercal{\boldsymbol\eta}+{\bf{B}}^{-1}{\bf{b}})$, $\boldsymbol\eta=(y_1-1/2,\ldots,y_n-1/2)^\intercal$ and $\boldsymbol\Omega=\mbox{diag}(\omega_{1},\ldots,\omega_{n})$. Efficient methods for sampling from P\'olya-gamma random variables are provided in \cite{Pol:2013}, and are implemented in the {\rm R} library \texttt{BayesLogit}.

We adapt the above Gibbs sampling algorithm to incorporate functional age effects via Bayesian penalized splines, and State-specific effects whose prior distribution is a mixture of Gaussians. This is accomplished by combining the P\'olya-gamma data augmentation  with classical Gibbs samplers for Bayesian finite mixtures of Gaussians, and for Bayesian penalized splines \citep{Lan:2004}. Leveraging the reparameterization of the multinomial likelihood in equation \eqref{repa}, the Gibbs algorithms to obtain posterior samples for the parameters in representation \eqref{linpred} can be performed separately, and with the same derivations for each logistic regression in the sequential formulation \eqref{seq_logi}. 

We outline below the detailed steps to update the prior distributions for the parameters associated with the {\bf usage} choice model. The Gibbs samplers for the {\bf reversibility}  and the {\bf method} choice models proceed in the same way, conditioning on the appropriate statistical units. In particular in the {\bf reversibility} choice model only women using a contraceptive method are considered to update the prior distributions. Similarly, the Gibbs sampler for the parameters in  the {\bf method} choice model leverages only information of statistical units using contraceptives but not being sterilized. Source codes for the three Gibbs samplers are available at \url{https://github.com/tommasorigon/India-SequentiaLogit}.

Let $z_{ij1}=\sum_{r=2}^4 y_{ijr}$ denote the binary indicator for the use of contraceptive methods, the Gibbs sampler for the parameters in the {\bf usage} choice model---corresponding to $k=1$---with priors \eqref{eq5}--\eqref{eq3}, proceeds according to the following steps.

\vspace{10pt}

\begin{addmargin}[1.3em]{0em}% 1em left, 2em right
{\em Step 1:} Update each P\'olya-gamma augmented data $\omega_{ij1}$ from the full conditional $\omega_{ij1} \mid - \sim \mbox{\small{PG}}\{1,\mu_{i1}+ f_1(\texttt{age}_{ij})+{\bf{x}}_{ij}^\intercal{\boldsymbol\beta_1 }\}$, for every $i=1,\ldots,33$ and $j=1, \ldots, n_i$.

\vspace{5pt}

\noindent {\em Step 2:} In updating the functional effect of the variable \texttt{AGE} on the probability of using or not a contraception method, the full conditional distribution for the vector of parameters $\boldsymbol\gamma_1$ is
$$
\boldsymbol\gamma_1 \mid - \sim \mbox{N}\{({\bf{H}}_1^\intercal \boldsymbol\Omega_1 {\bf{H}}_1+\lambda_1{\bf{D}})^{-1}{\bf{H}}_1^\intercal \boldsymbol\eta_{\gamma_1},({\bf{H}}_1^\intercal  \boldsymbol\Omega_1 {\bf{H}}_1+\lambda_1{\bf{D}})^{-1}\},
$$
where $\boldsymbol\eta_{\gamma_1}$ is a vector with entries $\eta_{ij\gamma_1}=z_{ij1}-1/2-\omega_{ij1}(\mu_{i1}+{\bf{x}}_{ij}^\intercal{\boldsymbol\beta_1 })$, $i=1,\ldots,33$, $j=1, \ldots, n_i$ whereas ${\bf{H}}_1$ is the B-splines design matrix having row entries $\mathcal{B}_m(\texttt{age}_{ij})$, $i=1,\ldots,33$, $j=1, \ldots, n_i$, for every column $m=1,\dots,M$. Finally, $\boldsymbol\Omega_1$ is a diagonal matrix with  entries $\omega_{ij1}$, $i=1,\ldots,33$, $j=1, \ldots, n_i$, on its diagonal.

\vspace{5pt}

\noindent {\em Step 3:} Update the parameter $\lambda_1$, controlling the smoothness for the functional effect of the   variable \texttt{AGE},  from its full conditional gamma random variable $\lambda_1 \mid - \sim \mbox{Ga} \{a_{\lambda}+\text{rank}({\bf D})/2, b_{\lambda} + \boldsymbol\gamma_1^\intercal{\bf D}\boldsymbol\gamma_1/2 \}$. Notice that $\bf D$ is not a full rank matrix, and therefore $\text{rank}({\bf D}) = M - 2$.

\vspace{5pt}

\noindent {\em Step 4:} Exploiting the P\`olya-gamma data augmentation scheme, the full conditional distribution for the vector of parameters $\boldsymbol\beta_1$ encoding the effect of the dummy covariates on the probability of using or not a contraceptive is 
$$\boldsymbol\beta_1 \mid - \sim \mbox{N}\{({\bf{X}}_1^\intercal  \boldsymbol\Omega_1 {\bf{X}}_1+{\bf{B}}^{-1})^{-1}({\bf{X}}_1^\intercal \boldsymbol\eta_{\beta_1}+{\bf{B}}^{-1}{\bf{b}}),({\bf{X}}_1^\intercal  \boldsymbol\Omega_1 {\bf{X}}_1+{\bf{B}}^{-1})^{-1}\},$$ where $\boldsymbol\eta_{\beta_1}$ is a vector with entries $\eta_{ij\beta_1}=z_{ij1}-1/2-\omega_{ij1}\{\mu_{i1}+f_1(\texttt{age}_{ij})\}$, $i=1,\ldots,33$, $j=1, \ldots, n_i$ whereas ${\bf{X}}_1$ is the corresponding design matrix having row entries ${\bf{x}}_{ij}^\intercal $. 

\vspace{5pt}

\noindent {\em Step 5:} To update the State-specific parameters under prior  \eqref{eq5}--\eqref{eq6}, we combine the P\`olya-gamma data augmentation with classical algorithms for Bayesian mixtures of Gaussians. In particular---leveraging the mixture representation in equation \eqref{eq5}---we first allocate each State $i=2,\dots,33$, to one of the $h=1, \ldots H$, clusters by sampling each group indicator $G_{i1}$, $i=2,\dots,33$ from the full conditional categorical random variable with probabilities
\begin{eqnarray*}
\mbox{pr}(G_{i1}=h \mid - )= \frac{\nu_{h1} \mbox{N}(\mu_{i1}; \bar{\mu}_{h1},\sigma_1^{2})}{\sum_{s=1}^{H}\nu_{s1} \mbox{N}(\mu_{i1}; \bar{\mu}_{s1},\sigma_1^{2})},  \quad h=1, \ldots, H.
\end{eqnarray*}

\noindent {\em Step 6:} Consistent with \eqref{eq5}, $(\mu_{i1} \mid G_{i1}=h) \sim \mbox{N}(\mu_{i1}; \bar{\mu}_{h1},\sigma_1^{2})$. Therefore, the full conditional of each State-specific parameter $\mu_{i1}$, given the above cluster assignments, is easily available as
$$ \mu_{i1} \mid - \sim \mbox{N}\left(\frac{\sum_{j=1}^{n_i} [z_{ij1}-1/2-\omega_{ij1}\{f_1(\texttt{age}_{ij})+{\bf{x}}_{ij}^\intercal{\boldsymbol\beta_1 }\}]+ \bar{\mu}_{G_{i1}1}/\sigma_1^{2}}{1/\sigma_1^{2}+\sum_{j=1}^{n_i}\omega_{ij1}},\frac{1}{1/\sigma_1^{2}+\sum_{j=1}^{n_i}\omega_{ij1}} \right),$$
independently for every $i=2, \ldots, 33$.
\vspace{5pt}

\noindent {\em Step 7:} Since $\sigma^2_1$ is shared among all the mixture components, and provided that $(\mu_{i1} \mid G_{i1}=h) \sim \mbox{N}(\mu_{i1}; \bar{\mu}_{h1},\sigma_1^{2})$, the full conditional for $\sigma^{-2}_1=\tau_1$ can be easily obtained via 
$$\tau_1 \mid - \sim \mbox{Ga}(a_{\tau_1}+32/2, b_{\tau_1}+\sum_{i=2}^{33} [\mu_{i1}-\bar{\mu}_{G_{i1}1}]^2/2)$$

\noindent {\em Step 8:} Exploiting again the result $(\mu_{i1} \mid G_{i1}=h) \sim \mbox{N}(\mu_{i1}; \bar{\mu}_{h1},\sigma_1^{2})$, the full conditional for $\bar{\mu}_{h1}$ is
$$\bar{\mu}_{h1} \mid - \sim \mbox{N}\left(\frac{\sum_{i: G_{i1}=h} \mu_{i1}/\sigma_1^{2}}{1/\sigma^2_{\mu 1}+\sum_{i: G_{i1}=h} 1/\sigma_1^{2}}, \frac{1}{1/\sigma^2_{\mu 1}+\sum_{i: G_{i1}=h} 1/\sigma_1^{2}}\right),$$
independently for every $h=1, \ldots, H$.

\vspace{5pt}

\noindent {\em Step 9:} Update the mixing probability vector $(\nu_{11}, \ldots, \nu_{H1})$ from its full conditional Dirichlet distribution $\mbox{Dirich}\{1/H+\sum_{i=2}^{33} \mbox{1}_{(1)}(G_{i1}), \ldots, 1/H+\sum_{i=2}^{33} \mbox{1}_{(H)}(G_{i1})\}$.

\end{addmargin}

\section{Socio-demographic determinants underlying the contraceptive choices in India}\label{sec_4}
Recalling our research interests, we apply the sequential logistic regressions outlined in representation \eqref{seq_logi} to the contraceptive preference data described in Section \ref{sec_11}. Our main goal is to flexibly estimate and interpret the covariates effects on the binary choices characterizing the sequential decision process in Figure \ref{F_decision}, and quantify  the uncertainty of our conclusions. Before discussing our findings in Section \ref{sec_42}, we first compare the model proposed in Section \ref{sec_2}  with alternative parametric specifications to assess whether the Bayesian semiparametric model outlined in Section \ref{sec_22} effectively improves inference in the motivating application considered. We also study performance in out-of-sample prediction of contraceptive preferences, and compare the results with state-of-the-art methods specifically developed for classification tasks, in order to assess to what extent our Bayesian semiparametric model induces a flexible representation of the socio-demographic factors underlying the contraceptive choices. These assessments are described in detail in Section  \ref{sec_41}.

In performing Bayesian inference we rely on the priors \eqref{eq5}--\eqref{eq3}, described in Section \ref{sec_22}. As there are at most $32$ clusters among the $32$ parameters $\mu_{2k}, \dots,\mu_{33k},$ characterizing the State-specific effects  in each logistic regression, we fix $H=32$, and allow the sparse Dirichlet prior in  \eqref{eq6} to delete redundant mixture components \citep{Rou:2011}. In setting $\sigma_{\mu k}^2$ and $(a_{\tau_k},b_{\tau_k})$ in \eqref{eq6}, note that $\sigma_{\mu k}^2$ measures the variability of the component-specific mean parameters with respect to zero, whereas $(a_{\tau_k},b_{\tau_k})$ controls the variance $\sigma^{2}_k$ of the State-specific effects in each mixture component. For instance, high values of  $\sigma_{\mu k}^2$ combined with a small $\sigma^{2}_k$, characterize situations in which States within the same cluster have highly similar State-specific effects, with these effects substantially changing  between the different clusters. Hence, in setting these hyperparameters, we consider a data-driven approach. Specifically, we first estimate a classical logistic regression with State-specific fixed effects, and then group these estimated effects via standard methods for clustering. Based on the resulting empirical clusters, $(a_{\tau_k},b_{\tau_k})$ and $\sigma_{\mu k}^2$ are specified using the information on the within cluster variance, and the averaged squared deviations of the cluster means from zero.  Consistent with this procedure we set $\sigma_{\mu 1}^{-2}=0.2$, $\sigma_{\mu 2}^{-2}=0.02$, $\sigma_{\mu 3}^{-2}=0.01$, and $(a_{\tau_1}=0.5,b_{\tau_1}=0.1)$, $(a_{\tau_2}=0.1,b_{\tau_2}=0.1)$, $(a_{\tau_3}=0.15,b_{\tau_3}=0.1)$. Although other settings are possible---using for example prior knowledge of inter-State differences in India---such empirical Bayes approach is typically useful in Bayesian hierarchical models to improve mixing and convergence. We also attempted inference under moderate changes of the above hyperparameters avoiding data-driven prior settings, but found no evident differences in the final results thanks to the borrowing of information induced by the mixture of Gaussians.

In defining the functional effect of the variable \texttt{AGE} in representation \eqref{eq:splines}, we follow instead \citet{Eil:1996} by relying on a sufficiently large number $n_\texttt{knots} = 46$ of equally spaced knots, and facilitate a moderate shrinkage by letting  $a_{\lambda}=1.5$ and $b_{\lambda}=5\cdot 10^{-4}$---consistent with the smoothness of the empirical trajectories in Figure \ref{predictor}. Finally, although the prior mean vector ${\bf{b}}$, and the covariance matrix ${\bf{B}}$, could be set according to current knowledge on the effects of the qualitative covariates, we let ${\bf{b}}={\bf{0}}$ and ${\bf{B}}=\mbox{diag}(100, \ldots, 100)$, to incorporate  the neutral hypothesis of no relevant effects, with a moderate prior uncertainty, since there is not overall agreement in current studies on $\boldsymbol\beta_k$. Also in this case we found posterior inference robust to moderate changes in ${\bf{b}}$ and ${\bf{B}}$. This is due to the fact that the  variables \texttt{AREA}, \texttt{RELIGION}, \texttt{EDUCATION}, and \texttt{CHILD} have a small number of well represented categories. Hence, there is sufficient information in the data to provide robust  inference on $\boldsymbol\beta_k$.

In performing posterior inference we consider $22000$ Gibbs iterations, holding out the first $2000$ as a burn-in, and thinning the chains every $5$ samples. As a result posterior inference relies on $4000$ Gibbs samples, for which the trace plots show no evidence against converge, and good mixing---monitored via effective sample sizes. Source code and step-by-step tutorials to reproduce the results discussed in Sections \ref{sec_41}--\ref{sec_42} are available online at  \url{https://github.com/tommasorigon/India-SequentiaLogit}. 

\vspace{-5pt}
\subsection{Model comparisons and out-of-sample predictive performance}\label{sec_41}
As discussed in Section \ref{sec_22}, an important contribution of the proposed statistical model is in improving flexibility compared to current analyses of contraceptive preferences in India. This increased flexibility is motivated by the data under analysis, and is accomplished by considering mixtures of Gaussians for the State-specific effects, along with penalized splines for the functional effects of the variabile   \texttt{AGE}. 

\subsubsection{Comparison with sub-models via DIC and WAIC}
To empirically assess the practical usefulness of the proposed formulation, we compare our  \textsc{mixture--splines} model with simpler sub-models, using the DIC and WAIC \citep{Gel:2014}. Consistent with the above considerations, and with the discussions in Section \ref{sec_22}, we consider three sub-models. These three alternative specifications comprise a \textsc{baseline} model in which the variable \texttt{AGE} enters the predictor linearly, and the parameters $\mu_{2k},\dots, \mu_{33k}$ have classical Gaussian priors; a \textsc{splines}  model replacing the linearity assumption in the \textsc{baseline}  model with the spline representation in \eqref{eq:splines}; and a \textsc{mixture}  model in which mixtures of Gaussians \eqref{eq5}--\eqref{eq6} are considered for  $\mu_{2k},\dots, \mu_{33k}$, but the variable \texttt{AGE} enters the predictor linearly as in the \textsc{baseline} model.

The above simpler sub-models are special and more parsimonious versions of our semiparametric specification, thereby proving relevant alternative representations to assess the actual usefulness of the increased flexibility provided by the  \textsc{mixture--splines} model. Posterior inference for these sub-models proceeds under minor modifications of the Gibbs sampler proposed in Section \ref{sec_3}, and the hyperparameters are set according to the same guidelines considered for our   \textsc{mixture--splines} model. Note also that, the simple factorization \eqref{repa} of the full-model likelihood, together with the independence of the priors distributions for $k=1,\dots,3$, allow to evaluate the partial DIC and WAIC for each logistic regression in \eqref{seq_logi}, and then obtain those for the full model by simple summation of the partial ones.

\begin{table}
\caption{DIC and WAIC information criteria for competing sub-models. The value $-2\times \text{WAIC}$ is reported to obtain indexes on the same scale. Bold values are the lowest (best) DIC and $-2\times \text{WAIC}$ respectively \label{tab:IC}}
\centering
\begin{tabular}{rllll}
  \toprule
&\ \ \ \ \ \ \ \ \ \ \ \ \ \ \ \ \  \textsc{baseline} &  \ \ \  \ \ \    \ \ \  \textsc{splines} &  \ \ \  \ \ \   \ \ \  \textsc{mixture}&  \ \ \   \ \ \ \ \ \  \textsc{mixture--splines}  \\ 
  \midrule
 DIC&\ \ \ \ \ \ \ \ \ \ \ \ \ \  \ \ \ 53507.70	& \ \ \  \ \ \  \ \ \  53092.90& \ \ \	 \ \ \  \ \ \  53505.00	& \ \ \  \ \ \  \ \ \  {\bf 53091.25} \\
 $-2\times \text{WAIC}$& \ \ \ \ \ \ \ \ \ \ \ \ \ \ \ \ \ 53503.41	& \ \ \    \ \ \  \ \ \ 53088.62	& \ \ \  \ \ \   \ \ \ 53498.06	& \ \ \  \ \ \  \ \ \ {\bf 53083.61}  \\ 
   \bottomrule
\end{tabular}
\end{table}

As shown in Table~\ref{tab:IC}, the DIC and WAIC indexes are on a similar scale. More evident improvements are obtained when modeling the functional effect of the variable \texttt{AGE} under a spline representation as in \eqref{eq:splines}, instead of a linear one. This supports our choice of improving flexibility in characterizing non-linear effects of the variable \texttt{AGE}. When comparing the \textsc{splines}  model with our  \textsc{mixture--splines} representation, we observe also an advantage in using mixtures of Gaussians instead of Gaussians priors for the State-specific effects, thereby confirming the usefulness of our semiparametric specification. This additional improvement is less evident compared to the introduction of non-linear effects for the variable \texttt{AGE}, meaning that a common Gaussian prior is reasonable for several States, but a subset of them may still present notable deviations from this assumption. Hence, incorporating this behavior in our model can provide key insights on relevant deviations for groups of States from shared structures.

\vspace{-7pt}
\subsubsection{Out-of-sample predictive performance}
The results  in Table~\ref{tab:IC} confirm that our Bayesian semiparametric formulation is empirically preferred over simpler specifications, but do not guarantee that the proposed model provides an accurate representation of the socio-demographic factors underlying the contraceptive preferences. For instance, although we improve flexibility via Bayesian splines and mixture modelling, the additive assumption for the effects of the different variables in equation \eqref{seq_logi} may still provide a restrictive representation of the determinants driving the contraceptive choices.

To understand whether the proposed statistical model is sufficiently flexible, we study the performance of our representation in out-of-sample prediction of the contraceptive preferences, and compare the results with those obtained under benchmark methods for classification---e.g. discriminant analysis, random forests, and gradient boosting---using the same covariates. These methods are  specifically developed to allow accurate predictions of a response variable leveraging much complex partitions of the covariates space. Therefore, a similar predictive performance under our model would provide relevant insights on the sufficient flexibility of the proposed representation, and its adequacy in accurately characterizing the  socio-demographic factors underlying the contraceptive preferences.

\begin{table}
\caption{Upper table: For our model and relevant competitors, out-of-sample performance in predicting the use or not use of contraceptive methods---measured via the AUC (area under the ROC curve), and the misclassification rate with cut-off $0.5$. Lower table: For our model and relevant competitors, out-of-sample performance in predicting use of natural methods, or modern methods or sterilization, for those women using contraceptives. \label{tab:pred}}
\centering
\begin{tabular}{lllllr}
  \toprule
 & \textsc{mixture--splines} &    \textsc{gradient boosting} & \textsc{random forest} &  \textsc{lda}& \textsc{multinom} \\ 
  \midrule
  \multicolumn{4}{l}{Predictive performance for \textbf{usage} choice }  \\ 
  \hline
  AUC & 0.799 & 0.803  & 0.797 & 0.790 &  --- \\ 
  Misclassification Rate & 0.197  & 0.194  &0.196 & 0.196 & --- \\ 
 \midrule
 \multicolumn{4}{l}{Combined predictive performance for \textbf{reversibility} and \textbf{method} choice }  \\ 
\hline
  Misclassification Rate  & 0.230  & 0.231 & 0.234 & 0.249 & 0.233 \\ 
   \bottomrule
\end{tabular}
\end{table}

As shown in Table \ref{tab:pred}, the predictive checks proceed with reference to two nested partitions of the response variable. The reason is that we additionally aim to compare predictive performance with a formulation recalling the one proposed by   \cite{Deo:2014}, who focus on comparing natural and modern methods against sterilization, only for women currently using contraceptives. Consistent with this, we first study the  predictive performance associated with the \textbf{usage} choice model alone---thereby focusing on the probability that a new individual will use contraceptive methods or not. Then, in the subsequent assessment, we study performance in predicting use of natural methods, modern methods or sterilization for the subset of women using contraceptives, consistent with   \cite{Deo:2014}. Both assessments are made on a subset of randomly selected women---comprising $25\%$ of the sample---using the remaining statistical units as training set for the different methods.

Out-of-sample predictions under our model formally rely on the expectation of the posterior predictive distribution of the contraceptive preference indicators, which coincide with the posterior mean of the conditional probabilities of interest. In particular, in the first assessment, we compute for every out-of-sample unit the posterior mean of the associated probability of using $\rho_{ij1}$ or not  using $1-\rho_{ij1}$ a contraceptive method, and then predict the final outcome  by checking if this estimated probability exceeds or not a specific cut-off. A similar procedure holds for the second assessment, except for focusing on the posterior mean of $(1-{\rho}_{ij2}), \rho_{ij2}{\rho}_{ij3}$ and ${\rho}_{ij2}{\rho}_{ij4}$, measuring the conditional probabilities of sterilization, modern and natural methods, respectively, given the decision to use contraceptives. In this case, the predicted category is the one having the highest estimated probability.

 As shown in Table~\ref{tab:pred}, although our statistical model is mainly focused on providing interpretable inference for the determinants underlying  the contraceptive preferences in India, we obtain a predictive performance in line with the benchmark methods specifically developed for prediction tasks. Moreover, the misclassification rate of our Bayesian semiparametric formulation is slightly lower than existing parametric multinomial models, such as the one proposed by \cite{Deo:2014}, thus suggesting that more flexible specifications are indeed preferred. In fact, in implementing a similar version of the multinomial regression in \cite{Deo:2014} we consider a piecewise constant specification for \texttt{AGE} in the intervals $[15,25]$, $(25,34]$ and $(34,49]$, which assumes that the effect of the variable \texttt{AGE} is the same within each interval.

 Although the above assessments are based on a simple hold-out approach, it is worth noticing that the accurate predictive performance is a side benefit of our statistical model, and the overarching focus is on providing meaningful and accurate inference. Hence, consistent with our fundamental goal, we avoid further complications via cross-validation and leverage the results in Table~\ref{tab:pred} to obtain simple reassurance that the proposed Bayesian semiparametric representation does not lead to  inadequate characterization of the socio-demographic factors underlying the contraceptive preferences. Moreover, we obtained similar conclusions when considering different training and test sets.

% This result provides a valuable insight on the accuracy of our model in flexibly characterizing the decision process associated with the observed data.

\subsection{Interpretation of the results}\label{sec_42}
The results in Section \ref{sec_41} confirm that the Bayesian semiparametric model proposed in Section \ref{sec_22} provides a sufficiently flexible and empirically motivated representation for the determinants underlying the sequential decision process discussed in Section \ref{sec_21}, with a specific reference to the current family planning programs in India. These results motivate discussion and interpretation of our findings via inference on the posterior distributions for the parameters in equation \eqref{seq_logi}.

The posterior distributions for the effects of variables \texttt{AREA}, \texttt{RELIGION}, \texttt{EDUCATION} and \texttt{CHILD}, are summarized in Table \ref{tab:2}, and provide interesting findings on the determinants of the contraceptive preferences. Lack of information about family planning, inaccessibility issues, marked son preference \citep[e.g.][]{Cha:2009} and the decreased woman empowerment \citep[e.g.][]{Lee:2010} in rural areas motivate a lower probability of contraceptive use---compared to urban areas---along with a reduced preference for reversible methods instead of sterilization. Consistent with this result, the preference for modern temporary methods---compared to natural ones---increases  in urban areas with respect to rural areas. This  is also in line with a reduced knowledge in rural areas about condoms and their additional effect in preventing sexually transmitted diseases \citep[e.g.][]{Don:2014}.

\begin{table}
\caption{Posterior mean and 0.95 credible intervals for the $\boldsymbol\beta_k$ parameters in the sequential logistic regressions. The bold parameter estimates have 0.95 credible intervals not including the value $0$. \label{tab:2}}
\centering
\begin{tabular}{lcrclrclrl}
  \toprule
  & & \multicolumn{2}{c}{\textbf{Usage}  } & & \multicolumn{2}{c}{\textbf{Reversibility}}&  &\multicolumn{2}{c}{\textbf{Method}}    \\\midrule
 \multicolumn{9}{l}{{\textsc{variable} \texttt{AREA}. Reference category: \texttt{rural}} }  \\   
 \hline
\texttt{urban} &$\quad \ \ \ \ $ &  ${\bf0.24}$ & \scriptsize{$[ \ 0.17 ; \ 0.31\ ]$}& & {\bf 0.29} & \scriptsize{$[ \ 0.21 ; \ 0.38\ ]$} & &  ${\bf 0.45}$ & \scriptsize{$[ \  0.32 ; \ 0.58 \ ]$} \\ 
 \midrule
\multicolumn{9}{l}{{\textsc{variable} \texttt{RELIGION}. Reference category: \texttt{hindu}} }  \\ 
 \hline
\texttt{muslim} &&  ${\bf-0.43}$ & \scriptsize{$[-0.52 ; -0.34]$} &$\quad $&  ${\bf1.24}$ & \scriptsize{$[\ 1.12 ;\  1.36\ ]$} &$\quad $ &   $0.13$ & \scriptsize{$[ -0.03 ; \ 0.29 \ ]$}\\ 
\texttt{christian} &&  ${\bf-0.26}$ & \scriptsize{$[-0.48 ; -0.03]$} && $0.00$ &\scriptsize{$[-0.29 ; \ 0.29 \ ]$} & &  $0.36$ & \scriptsize{$[-0.13 ; \ 0.88 \ ]$} \\ 
\texttt{other} &  &$0.08$ & \scriptsize{$[-0.11 ; \ 0.29 \ ]$} && ${\bf0.46}$ & \scriptsize{$[\ 0.25 ; \ 0.66 \ ]$} & &   ${\bf0.30}$ & \scriptsize{$[\ 0.02 ; \ 0.59\  ]$}\\ 
 \midrule
  \multicolumn{9}{l}{{\textsc{variable} \texttt{EDUCATION}. Reference category: \texttt{no education}} }  \\ 
 \hline
\texttt{low} & & ${\bf0.14}$ & \scriptsize{$[ \ 0.05 ; \ 0.23 \ ]$} & & $0.08$ & \scriptsize{$[-0.03 ; \ 0.19 \ ]$} & &   ${\bf0.43}$ & \scriptsize{$[\ 0.25 ; \ 0.60 \ ]$} \\ 
\texttt{intermediate} & & ${\bf0.20}$ & \scriptsize{$[ \ 0.12 ; \ 0.27 \ ]$} && ${\bf0.50}$ & \scriptsize{$[\ 0.40 ; \ 0.59 \ ]$} & & ${\bf0.71}$ & \scriptsize{$[\  0.56 ; \ 0.86\ ]$} \\ 
\texttt{high} & & ${\bf0.27}$ & \scriptsize{$[\ 0.16 ; \ 0.37\ ]$} && ${\bf 1.28}$ & \scriptsize{$[\ 1.14 ;  \ 1.41 \ ]$} & &   ${\bf1.16}$ & \scriptsize{$[ \ 0.97 ; \ 1.34 \ ]$} \\ 
\midrule
 \multicolumn{9}{l}{{\textsc{variable} \texttt{CHILD}. Reference category: \texttt{more than one child}} }  \\  
 \hline
\texttt{no child} && ${\bf -3.71}$ & \scriptsize{$[-3.88 ; -3.54]$}& &  ${\bf2.20}$ &  \scriptsize{$[\ 1.69 ; \ 2.75 \ ]$}  & &   $-0.19$ &  \scriptsize{$[ \ -0.61 ;  \ 0.22 \ ]$}   \\ 
\texttt{one child} & & ${\bf-1.37}$ & \scriptsize{$[-1.45 ; -1.28]$} &&  ${\bf2.19}$ & \scriptsize{$[\ 2.06 ;  \ 2.32 \ ]$} & &   ${\bf-0.19}$ & \scriptsize{$[ -0.34 ; -0.04 ]$} \\ 
   \bottomrule
\end{tabular}
\end{table}

Focusing on the religion effect, there is a literature providing comparisons between muslims and hindus with respect to contraceptive behavior \citep[e.g.][]{Dha:2004}. The substantially different fertility intentions between these two religions, motivates a reduced use of contraceptives for muslims---compared to hindus. Additionally, Islam opposition to sterilization is evident in the reversibility choice with an increased preference for temporary contraceptives than hindus---among partners using contraceptive methods.  Christians are instead more similar to hindus, with exception of a reduced  attitude towards the use of contraceptives. This result is in line with the lack of formal prohibitions with respect to contraceptions in hinduism, making this practice acceptable. Refer also to \citet{Sri:2008} for an additional discussion of religion influences on contraception.

Consistent with recent contributions \citep[e.g.][]{Riz:2012}, growing literacy has an increasing positive effect in the decision of using contraceptives and in avoiding sterilization---among partners considering contraceptive methods. These effects are reasonably favored by higher accessibility, better knowledge of contraception, lower son preference and increasing possibility for women  empowerment within the household. Highly educated individuals are further characterized by an increased preference for modern methods---among partners opting for temporary contraceptives. This finding is in line with an increased awareness on sexually transmitted diseases \citep[e.g.][]{Don:2014}.

Finally---compared to women having more than one child---we observe an increasingly lower preference towards contraceptive use and sterilization in the sequential process characterizing women with no or one child. These results are in line with recent findings, and are associated with the different fertility intentions and son preferences at varying number of children \citep[e.g.][]{Das:1986}.

\begin{figure}[t]
\centering
\includegraphics[height=5.5cm, width=17cm]{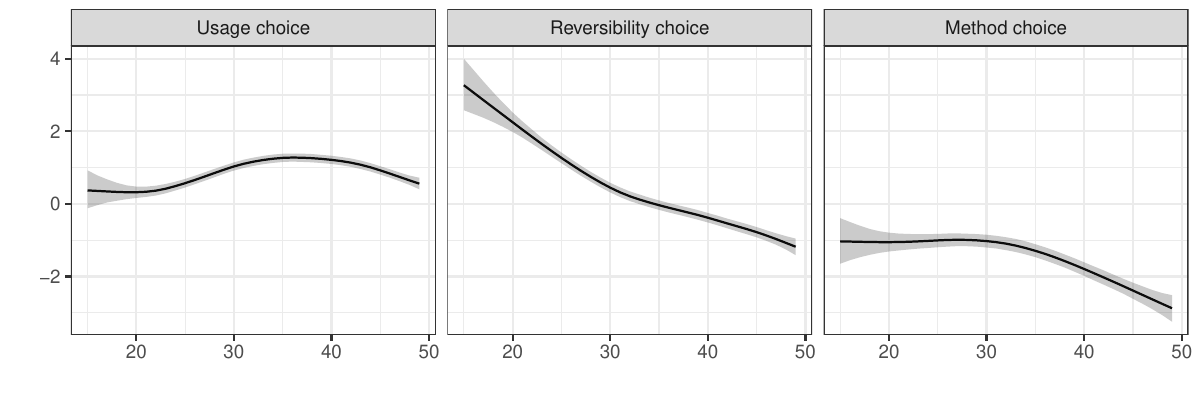}
\caption{\footnotesize{Posterior mean and point-wise 0.95 credible intervals (gray areas) for the functional effect of the variable \texttt{AGE} in each sequential logistic regression.}\label{predictor1}}
\end{figure}

\begin{figure}[h!]

\begin{minipage}{\textwidth}
\centering
\includegraphics[height=16cm,width=14cm]{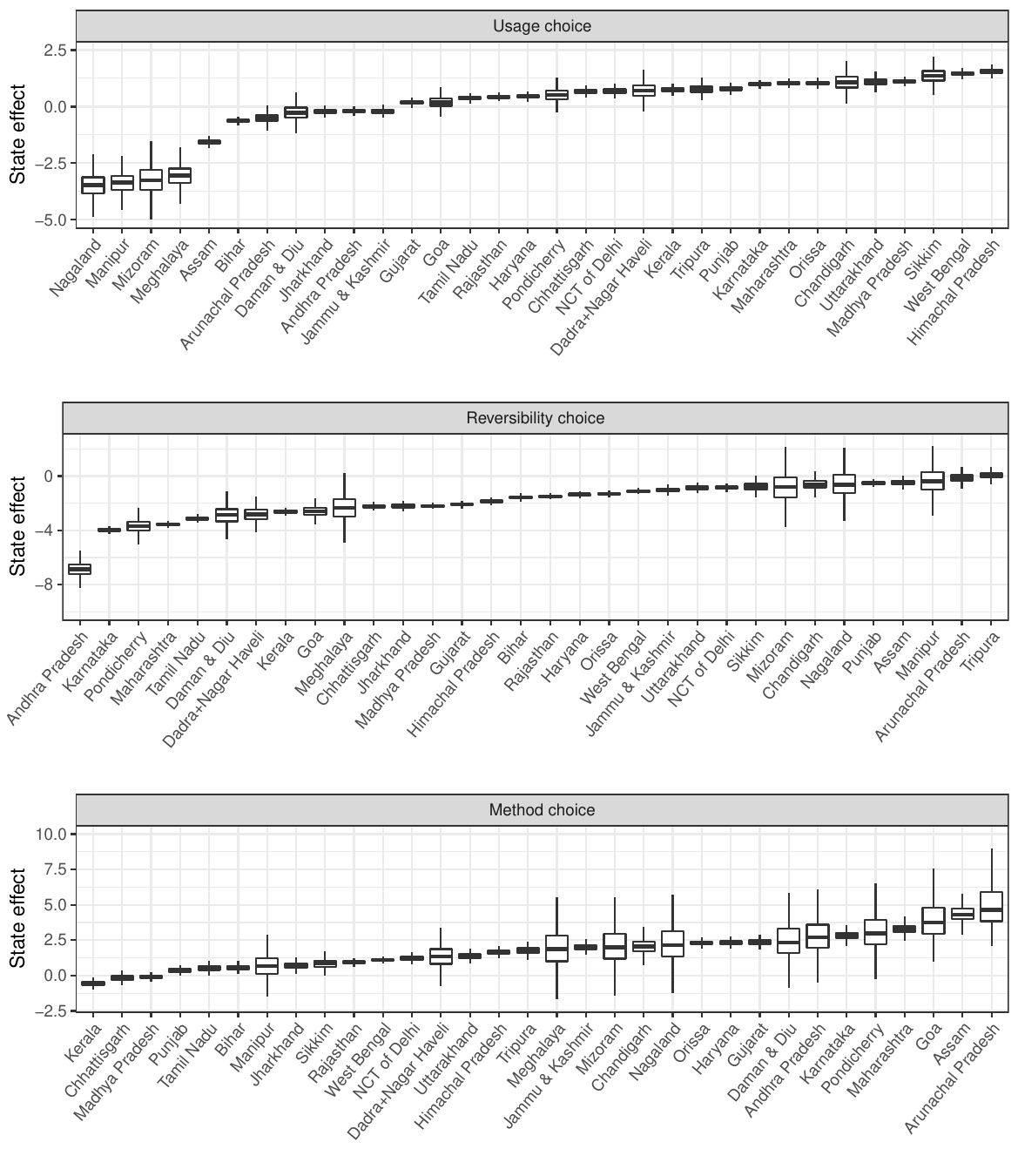}
\end{minipage}

\begin{minipage}{\textwidth}
\centering
\includegraphics[height=6cm, width=14cm]{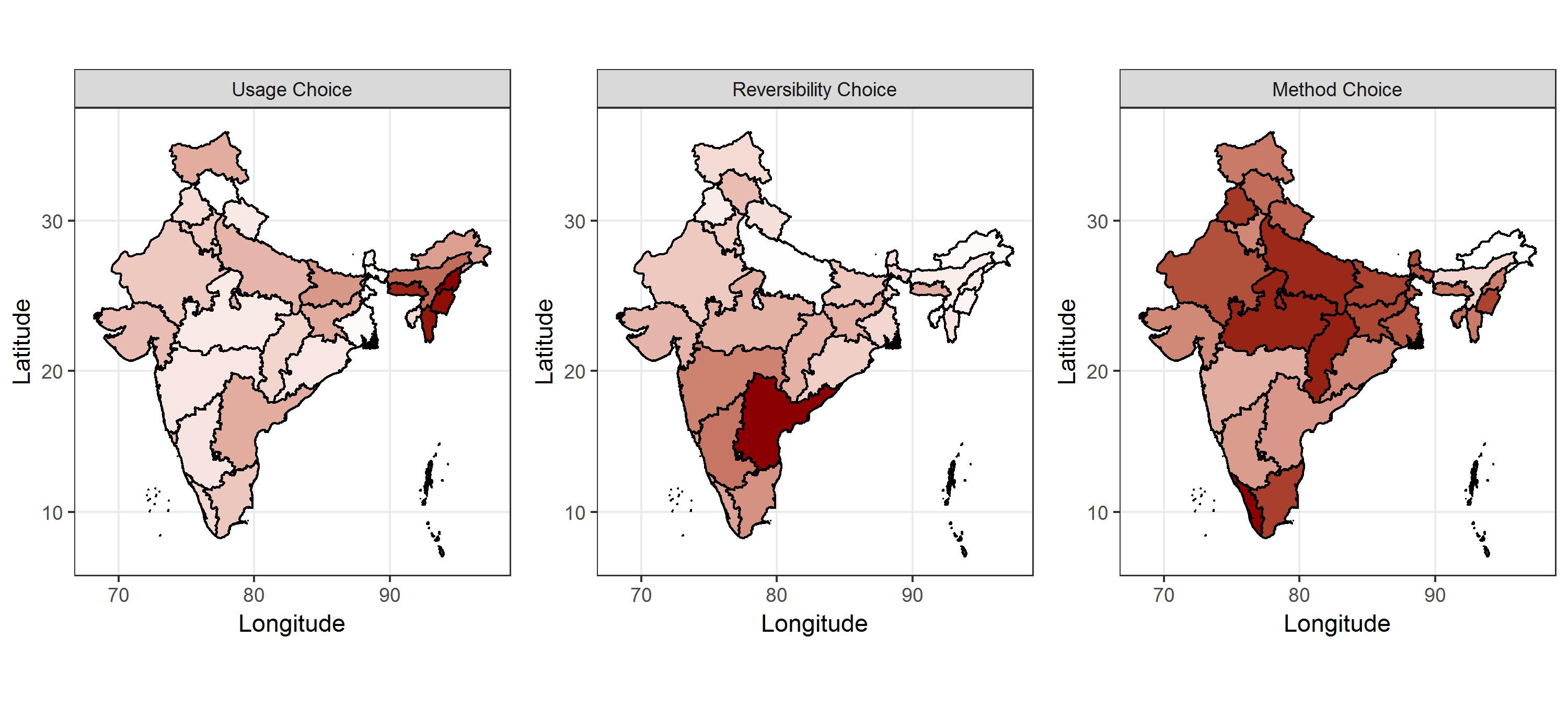}
\end{minipage}

\caption{\footnotesize{Upper panels: boxplots summarizing the posterior distribution of the State-specific effects in each sequential logit. Lower panels: States of India colored according to their corresponding estimated effect in each logist regression for the different steps of the sequential decision process in Figure~\ref{F_decision}.}\label{map}}
\end{figure}

Figure \ref{predictor1} summarizes the posterior distribution for the functional effect of the variable age at the different steps of the decision process underlying the contraceptive preferences. As expected the effect on the contraceptive use has an overall parabolic trend, peaking between 30 -- 40 years when birth control is more common. Sterilization is still the most common contraceptive method in India  \citep[e.g.][]{Pac:2004, Pac:2014} and its prevalence tends to increase with age \citep[e.g.][]{sav:1999}, motivating the previously discussed trend along with the decreasing functional effect of the variable age on the choice of temporary methods instead of sterilization. Among the women opting for reversible methods, we observe an increasing preference towards natural strategies---compared to modern methods---as age grows. This trend is evident only after 30 years, meaning that young couples have still, potentially unmet, interest towards modern methods  \citep[e.g.][]{Pac:2004}.

The above results are mostly in line with recent studies on the contraceptive preferences in India \citep[e.g.][]{Mcn:2003, Hus:2013, Rai:2013, Deo:2014,  DeoDias:2014, Ram:2014,  Mis:2014, Haq:2015}. However, as already discussed in Sections \ref{sec_1} and \ref{sec_2}, our model provides a more global and flexible overview which is motivated by current family planning policies in India, and avoids focusing only on specific aspects of the contraceptive preferences in India.

To conclude our study, we focus on the posterior distributions of the State-specific incremental effects with respect to the baseline Uttar Pradesh. Socio-economic differences across States---or groups of States---are marked in India \citep[e.g][]{Bal:2010, Das:1999}. Current studies exploring State-specific differences in contraceptive preferences typically focus on sub-samples of States \citep[e.g.][]{Rai:2013}, or groups of States \citep[e.g.][]{Deo:2014,  Ram:2014}, or estimate different models for each State \citep[e.g.][]{Dha:2004}. \citet{Dha:2004} consider also a multilevel analysis, but their focus is on religion differences in contraceptive preferences. Our model allows instead inference also on the State-specific effects after controlling for other covariates.

The benefits associated with our mixture of Gaussians prior \eqref{eq5}--\eqref{eq6} for the State-specific parameters are clear in Figure~\ref{map}, which shows groups of States whose effects are not forced to over-shrink around the global mean, and displays clustering effects that are interestingly in line with the geographical positions of the different States---without informing the model of such geographical structure. This improved flexibility highlights a substantially lower intention towards contraceptive use for a group of North Eastern States including Nagaland, Manipur, Mizoram, and Meghalaya. This clustering tendency for the North Eastern States is also evident in the logistic regression associated with the reversibility choice, and is further confirmed when applying the procedure of \citet{Med:2002} for clustering in mixture models. This result is in line with the common political history and specific cultural aspects of the  North Eastern States  in India, which are also referred to as ``seven sisters"  \citep{Bar:2006}. Andhra Pradesh displays instead a substantially lower preference for temporary methods, compared to sterilization. This confirms the strong measures adopted by the government of Andhra Pradesh to promote sterilization \citep{Pra:2009}. 

Finally, it is also worth noticing that in  Figure~\ref{map} some boxplots are wider than others, due to the fact that some States have reduced information for specific contraceptive preferences. This result further motivates our choice to improve borrowing of information via a Bayesian mixture of Gaussians for the State-specific effects, which still maintains flexibility in modelling more evident deviations.

\section{Conclusion}\label{sec_5}
Contraceptive preferences are subject to a complex combination of family planning policies and  socio-demographic differences in India, thereby requiring meaningful statistical models and flexible inference procedures to provide interpretable and accurate conclusions. The available statistical models are not sufficiently flexible, and typically fail to provide a global overview of the determinants underlying the entire decision process characterizing the contraceptive choices. To address this gap, we developed a Bayesian semiparametric statistical model relying on a set of logistic regressions which characterize a sequential decision process motivated by the current family planning policies in India. 

Our results substantially agree with the descriptive analyses available from other national surveys such as NFHS, and are typically in line with findings from other contributions studying only a subset of the entire decision process  in Figure \ref{F_decision}. A major benefit of our formulation is in allowing inference and uncertainty quantification on the entire set of contraceptive preferences, disambiguating the effects of the socio-demographic covariates at every step of the sequential process in Figure \ref{F_decision}, within a unique and flexible statistical model. This approach to inference provides a global and interpretable overview of the entire contraceptive preferences, facilitating the assessment of current family planning policies, and an improved targeting of subpopulations not yet addressed. 

%Data from IHDS--II (or NFHS) are collected in surveys with complex design. A possible concern in these cases is about taking into account surevey weigths. In Bayesian data analysis has been generally accepted that models should include (condition on) all variables that affect the probability of inclusion and non response. This issue has been  reviewed in the literature on Bayesian modeling \citep[e.g.]{Gel:2007} and recently possible solutions have been proposed in standard and simpler models \citep{Sie:2015}. In our case, we model the response conditionally on the variables considered in the survey design. Moreover available survey weigths for IHDS--II did not take into account calibration for non--response and other non sampling errors. Finally, evaluation of possible effects of alternative strategies for taking into account survey weigths in Bayesian non parametric models is a topic well beyond the scope of this paper.

Although alternative decision mechanisms could be considered, the process in Figure \ref{F_decision} is of interest as discussed in Section \ref{sec_21}, and provides  a formal reparameterization of the multinomial probability mass function for the contraceptive preferences. This facilitates also posterior inference on the contraceptive probabilities in equation \eqref{eq1.1} for every configuration of covariates characterizing the different women profiles. These quantities of interest, along with the conditional probabilities of the sequential choices in Figure \ref{F_decision}, can be interactively calculated and visualized in a {\bf Shiny} application available at \url{https://github.com/tommasorigon/India-SequentiaLogit}.

\bibliographystyle{rss}

\end{document}